\documentclass[a4paper,twocolumn,english,pre,nofootinbib, superscriptaddress]{revtex4-2}
\usepackage[T1]{fontenc}
\usepackage{color}
\usepackage{babel}
\usepackage{mathtools}
\usepackage{amsmath}
\usepackage{amssymb}
\usepackage{graphicx}
\usepackage{enumerate}
\usepackage{enumitem}
\usepackage{hyperref}
\hypersetup{
        breaklinks=true,
	colorlinks=true,
	linkcolor=blue,
	citecolor=blue,
	urlcolor=blue
}
\graphicspath{{./Figures/}}

\DeclareMathOperator{\sign}{sign}

\begin{document}

\preprint{AIP/123-QED}

\title[Rare Events in Extreme Value Statistics of Jump Processes with Power Tails]{Rare Events in Extreme Value Statistics of Jump Processes with Power Tails}

\author{Alberto Bassanoni}
\email{alberto.bassanoni@unipr.it}

\affiliation{Dipartimento di Scienze Matematiche, Fisiche ed Informatiche,
Università degli Studi di Parma, Parco Area delle Scienze 7/A, 43124, 
Parma, Italy}

\affiliation{INFN, Gruppo Collegato di Parma, Università degli Studi di Parma, Parco Area delle Scienze 7/A, 43124, Parma, Italy}

\author{Alessandro Vezzani}
\email{alessandro.vezzani@unipr.it}

\affiliation{Dipartimento di Scienze Matematiche, Fisiche ed Informatiche,
Università degli Studi di Parma, Parco Area delle Scienze 7/A, 43124, 
Parma, Italy}

\affiliation{Istituto dei Materiali per l'Elettronica ed il Magnetismo (IMEM-CNR),Università degli Studi di Parma, Parco Area delle Scienze 37/A, 43124, Parma, Italy}

\author{Raffaella Burioni}
\email{raffaella.burioni@unipr.it}

\affiliation{Dipartimento di Scienze Matematiche, Fisiche ed Informatiche,
Università degli Studi di Parma, Parco Area delle Scienze 7/A, 43124, 
Parma, Italy}

\affiliation{INFN, Gruppo Collegato di Parma, Università degli Studi di Parma, Parco Area delle Scienze 7/A, 43124, Parma, Italy}

\date{August 30, 2024}

\begin{abstract}
We study rare events in the extreme value statistics of stochastic symmetric jump processes with power tails in the distributions of the jumps, using the big-jump principle. The principle states that in the presence of stochastic processes with power tails statistics, if at a certain time a physical quantity takes on a value much larger than its typical value, this large fluctuation is realised through a single macroscopic jump that exceeds the typical scale of the process by several orders of magnitude. In particular, our estimation focuses on the asymptotic behaviour of the tail of the probability distribution of maxima, a fundamental quantity in a wide class of stochastic models used in chemistry to estimate reaction thresholds, in climatology for earthquake risk assessment, in finance for portfolio management, and in ecology for the collective behaviour of species. We determine the analytical form of the probability distribution of rare events in the extreme value statistics of three jump processes with power tails; L\'evy flights, L\'evy walks and the L\'evy-Lorentz gas. For the L\'evy flights, we re-obtain through the big-jump approach recent analytical results, extending their validity. For the L\'evy-Lorentz gas we show that the topology of the disordered lattice along which the walker moves induces memory effects in its dynamics, which influences the extreme value statistics. Our results are confirmed by extensive numerical simulations.
\end{abstract}

\maketitle

\begin{quotation}
\bf{Extreme value statistics plays a crucial role in environmental science, finance, engineering and risk assessment. By studying the behaviour of extreme events, such as unusually large or small values in a set of random variables, one can gain insight into rare but potentially relevant events. If the random variables follow a fat-tailed probability distribution, rare events in extreme value statistics can become more likely. We study these rare events in the extreme values of three well-known one-dimensional jump processes that involve a power-tailed probability distribution, where the main contribution to the rare fluctuations is due to a single large event. We determine the analytical form of the probability distribution of the rare events and show that when the distributions are power-tailed, even if the typical scales are Gaussian, rare events induced by a single large jump give a major contribution to the statistics.}
\end{quotation}

\section{Introduction}
\label{introduction}

\noindent Extreme Value Statistics (EVS) studies the probability distribution of the maximum value in a set of random variables, a topic of great interest and fundamental to understanding the impact of rare events on many systems. Examples of cases where EVS is key include chemical reactions \cite{Gardiner}, earthquake risk assessment \cite{Matthews}, financial processes \cite{BouchaudFinance}, and natural and social phenomena \cite{AlbeverioNature, KantzSocial}.
\\
\noindent EVS becomes particularly interesting when dealing with variables drawn from a fat-tailed probability distribution. Extensive studies have been carried out on the EVS of finite or countable sets of independent and identically distributed (IID) random variables \cite{BouchaudUnivClassEVS,MajumdarReviewEVS}, as well as for stochastic and correlated transport processes with discrete sampling \cite{GuilletEVSTransport, EliDiscreteSamplingEVS}, for biased random walks \cite{ArtusoEVSBiasedRW} and, 
interestingly, also for the limit laws of max-min and min-max functions of random matrices \cite{MetzlerMaxMin}. Recent research has investigated the EVS of discrete-time and continuous-space one-dimensional jump processes with steps drawn from a continuous and symmetric probability distribution \cite{Klinger2023}. The behaviour of the running maximum in these jump processes has been found to be characterised by its expected value, which in its leading asymptotic depends on the tails of the probability distribution of each individual step \cite{Mounaix_2018, Bruyne_2021}.
\\
The big-jump principle is central to the study of rare events in this class of fat-tailed stochastic processes. The principle states that rare events are not the result of the cumulative effect of many small rare subevents, but are caused by the macroscopic contribution of a single rare event, called the big jump. This principle has been rigorously demonstrated for finite sets of IID variables \cite{Chistyakov, BJFoss} and has been proven for various fat-tailed processes, including those with L\'evy statistics \cite{BJ1}, such as L\'evy flights, L\'evy walks \cite{LevyZaburdaev}, generalised L\'evy processes with memory effects \cite{BJ2}, Weibull subexponential processes \cite{BJ_sub_exp, BJ_subexp_Hamdi}, and disordered lattice processes such as the L\'evy-Lorentz gas \cite{LevyLorentz1, BJ3, LevyLorentz2Univ}.
\\
The aim of this paper is to study the rare events in EVS of three well-known one-dimensional symmetric jump processes with power-tailed probability distributions (for some results on rare events of asymmetric processes see \cite{BurioniAsymmetric1, BurioniAsymmetric2}): L\'evy flights in section \ref{section2}, L\'evy walks in section \ref{section3}, and L\'evy-Lorentz gases in section \ref{section4}, using the big-jump principle. The aim is to calculate for each of them the probability that the running maximum will reach a threshold $X$ that is significantly larger than the typical scaling length of the process. According to the principle, the probability that the maximum is greater than $X$ can be expressed as the product of the probability that a single jump is greater than $X$ and the number of jump attempts made during a process of duration $T$. Therefore this probability can be more easily estimated by calculating these two terms separately, using explicit knowledge of the process contributing to the tail. 
Our results show that for the L\'evy-Lorentz gas, the topology of the disordered lattice induces memory effects in the dynamics of rare events. For L\'evy flights and L\'evy walks the motion of the walker after the big jump can indeed be neglected and one can show that the contribution of the big jump to the simple probability density of the position  is equal to that of the extreme, and the two tails coincide. On the other hand, for the L\'evy-Lorentz gas during the big jump, the walker is reflected in a big quenched gap. This motion cannot be neglected and gives different tails for the position and for the maximum.
\\
For the L\'evy flight, when the mean square length of a single step is infinite, our analysis reproduces the recent results obtained in \cite{Klinger2023, MajumdarRecords} in the continuum limit with fractional diffusion (see Appendix \ref{appendixA} for details). Interestingly, our approach also applies to scenarios where the jump length distributions are power-tailed but the mean square length of each jump is finite and a Gaussian diffusion emerges, highlighting the relevance of rare events induced by a single large jump even in these apparently regular systems. All predictions in the three cases are compared with numerical simulations, with excellent agreement. Our results underline the role of the single big-jump principle for the estimation of rare events in EVS and record statistics, extending its application beyond the calculation of the probability densities for the position at a given time of a random walker and, according to recent studies, also for its exit times \cite{BJexit}.

\section{Extreme Value Statistics of L\'evy Flights}
\label{section2}

\noindent One-dimensional L\'evy flights are discrete-time random walks where at each step the walker has a probability $1/2$ of moving left or right, and the length $r$ of the step is drawn from a power-tailed probability density function (PDF) $p(r)$. We consider a PDF $p(r)$ which has the following power law with a lower bound $r_0$:
\begin{equation}
\label{levyflight}
p(r) =
\begin{cases}
\frac{\alpha r_0^{\alpha}}{r^{1+\alpha}} & r>r_0 \\
0 & r<r_0
\end{cases}
\end{equation}

\noindent For $\alpha > 2$ the second moment of the distribution $p(r)$ is finite, implying standard diffusion, while for $\alpha < 2$ the second moment of the distribution diverges, implying superdiffusive behaviour. This result for L\'evy flights can be generalised to any probability distribution that has the same asymptotic behaviour for large $r$. Now consider a one-dimensional L\'evy flight and denote the position of the walker after $n$ steps as $x(n)$, assuming that it starts from the origin $x(0)=0$. This jump process is governed by the Markov rule:

\begin{equation}
\label{markov}
x(n+1) = x(n) \pm r(n)
\end{equation}

\noindent where the sign is drawn with probability $1/2$ and the increment $r(n)$ follows the probability distribution \eqref{levyflight}. The object of interest is $P(X, n)$, which is the probability density that the jump process reaches a maximum $X$ within $n$ steps. From the L\'evy-Gauss Central Limit Theorem, the bulk of the running maximum PDF converges in probability in the large $n$ limit to a scaling function $\ell^{-1}(n) f(X/\ell(n))$ where $\ell(n)$ is the typical length scale of the process.
In particular, $P(X,n) \sim \ell^{-1}(n) f(X/\ell(n))$ if $X < \ell(n)\kappa(n)$, where $\kappa(n)$ is a slowly growing function in $n$ (e. g. a logarithmic function). For a subexponential distribution $p(r)$ (as defined in \cite{BJFoss}), if $X$ is much larger than the typical length scale $\ell(n)$, i.e. $X>\ell(n)\kappa(n)$, the probability distribution of the running maximum $P(X, n)$ follows a different behaviour, and it can be estimated using the single big-jump principle \cite{Chistyakov}. In particular, $\ell(n) \sim n^{1/\alpha}$ for $\alpha<2$, i.e. L\'evy flights in the strict sense, and $\ell(n) \sim n^{1/2}$ for $\alpha>2$, i.e. diffusive processes. The principle states that  in a time $n$ a distance $X > \ell(n)\kappa(n)$ is reached by a single big jump of order $X$, and the other steps can be neglected. On the other hand, smaller distances of order $\ell(n)$ are typically reached in several steps. In this framework, it can be assumed that the only significant contribution that leads the running maximum of the jump process beyond the $X$ in $n$ steps is made by a single macroscopic jump of size larger than $X$. In this interpretation, the random walker has exactly $n$ trials to make this single big jump that takes the process beyond $X$. Thus the probability that the running maximum is greater than $X$ in the limit $X \gg \ell(n)$ is given by:

\begin{figure}
\includegraphics[scale = 0.6]{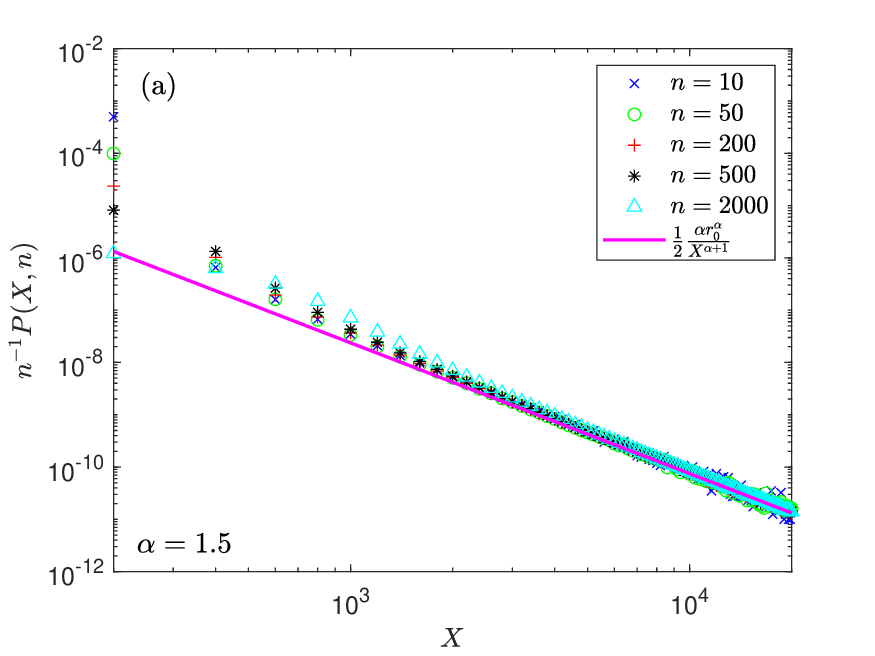}
\includegraphics[scale = 0.6]{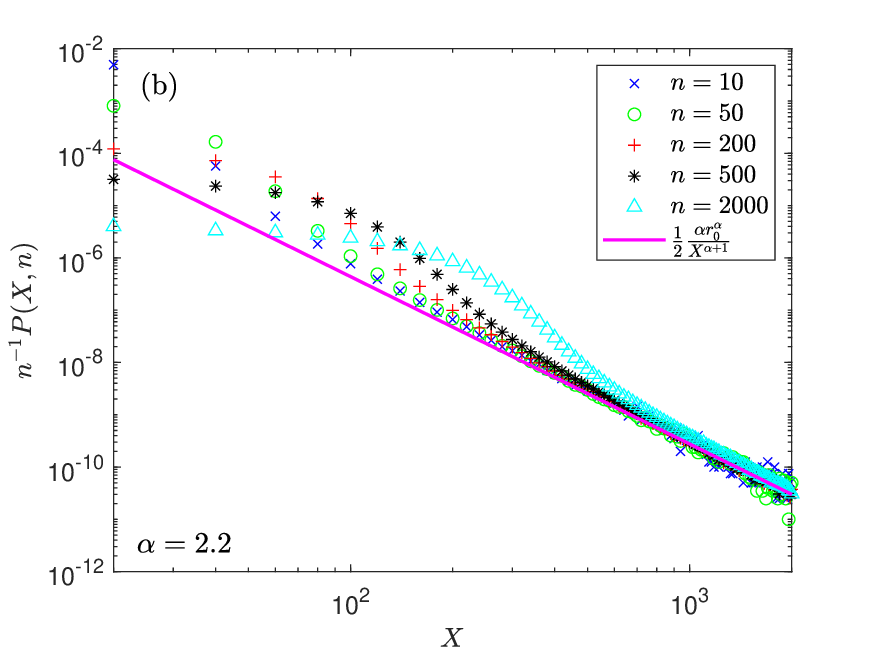}
\caption{Probability density function of running maximum $P(X, n)$ in logarithmic scale for L\'evy flights at several steps $n$. (a) plots an example of a superdiffusive walker with $\ell(n) \sim n^{1/\alpha}$ for $\alpha = 1.5$, while (b) plots an example of a diffusive walker with $\ell(n) \sim n^{1/2}$ for $\alpha=2.2$. In both simulations, $r_0=1$ was fixed and the probabilities were multiplied by a normalization factor $n^{-1}$. The continuous pink line represents the asymptotic analytic prediction found in Eq. \eqref{levyflightpdf};}
\label{fig:flight}
\end{figure}

\begin{equation}
\label{probflight}
\text{Prob}\left(\max_{i=1, ..., n} x(i) > X \right) \sim \frac{1}{2}  n \int_X^{\infty} dr \ p(r)
\end{equation}

\noindent where the factor $1/2$ accounts for the fact that every single increment of the jump process can be made in the positive or negative direction with probability $1/2$ and we exploit the fact that $\int_X^{\infty} dr \ p(r)$ is small for large $X$. Now, taking the derivative in $X$ of equation \eqref{probflight}, we get the PDF $P(X,n)$ of the running maximum:

\begin{equation}
\label{levyflightpdf}
P(X, n) \sim \frac{1}{2} n \frac{ \alpha r_0^{\alpha}}{X^{\alpha + 1}}
\end{equation}

\noindent The analytical prediction of Eq. \eqref{levyflightpdf} has been compared with the numerical simulations in Figure \ref{fig:flight} and shows excellent agreement in the regime of large $X$, both in the case of superdiffusive walkers for $\alpha<2$ and for standard diffusive walkers for $\alpha>2$. For $\alpha<2$ Eq. \eqref{levyflightpdf} confirms the results on EVS in the tail limit of L\'evy jump processes obtained in \cite{Klinger2023} with an approach based on survival probabilities analysis in the continuous limit. This result is extended also to the Gaussian case for $\alpha>2$, where the rare events at large distances are sub-leading with respect to the typical behaviour and  cannot be calculated by using the continuous limit. In particular, the power law decay \eqref{levyflightpdf} is still visible at any finite time $n$ and captures the behaviour of the extreme value distribution in the region $X > \ell(n)\kappa(n)$. The tail becomes relevant for all momenta $\langle X^q(n) \rangle$ with $q > \alpha$, causing divergences (for more details see \cite{BJ_sub_exp}).

\section{Extreme Value Statistics of L\'evy Walks}
\label{section3}

\begin{figure}
\includegraphics[scale = 0.6]{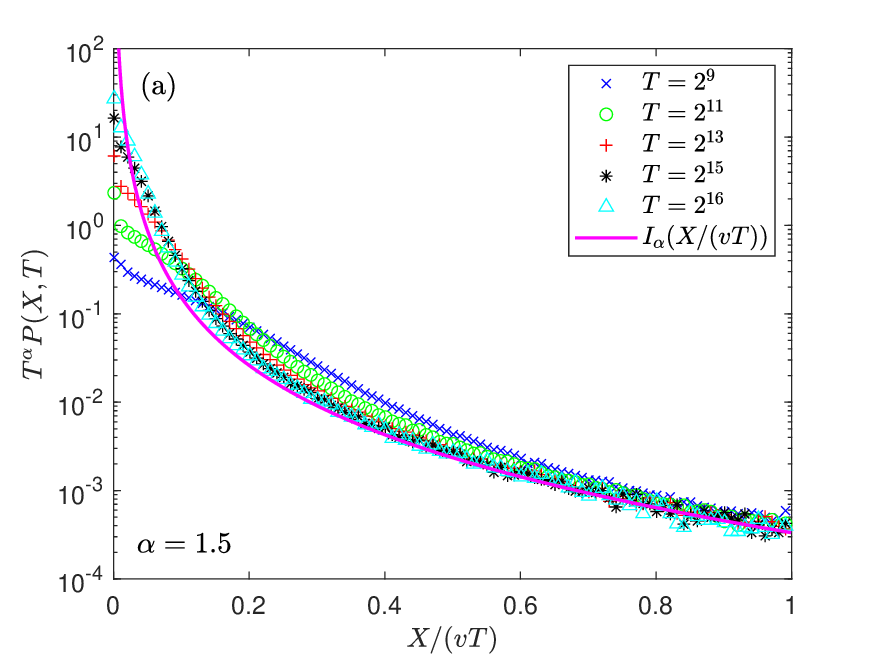}
\includegraphics[scale = 0.6]{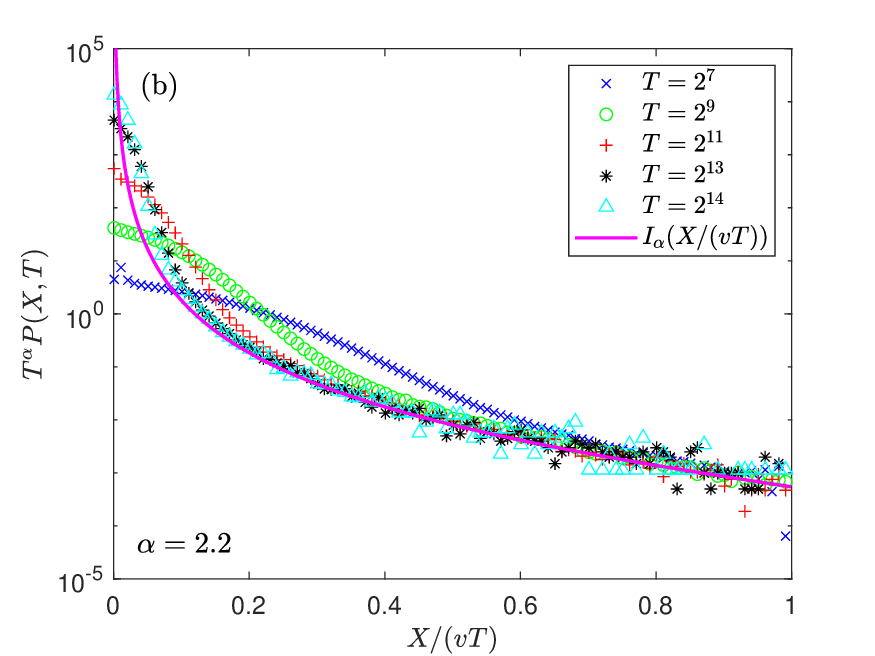}
\caption{Probability density function of running maximum $P(X, T)$ for L\'evy walks at different times $T$. (a) plots an example of a superdiffusive walker with $\ell(T) \sim T^{1/\alpha}$ for $\alpha = 1.5$, while Figure (b) plots an example of a standard diffusive walker with $\ell(T) \sim T^{1/2}$ for $\alpha=2.2$. In both simulations, $t_0=1, v=1$ were fixed and the probabilities were multiplied by a normalization factor $T^{\alpha}$. The continuous pink line represents the asymptotic analytic prediction in terms of infinite density function found in Eq. \eqref{levywalkpdf};}
\label{fig:walk}
\end{figure}

\noindent One-dimensional L\'evy walks are a special class of continuous-time random walks in which each step of the random walker has a time $t$ drawn from a probability distribution $p(t)$ with a power tail, and each step has a finite velocity $v_i$. Each individual step $r_i = v_i t$ is walked with probability $1/2$ at velocity $v_i = v$ and with probability $1/2$ at velocity $v_i=-v$ ($v>0$). The PDF of the time duration of the steps is defined as before by introducing a lower cut-off  $t_0$:

\begin{equation}
\label{levywalk}
p(t) =
\begin{cases}
\frac{\alpha t_0^{\alpha}}{t^{1+\alpha}} & t>t_0 \\
0 & t<t_0
\end{cases}
\end{equation}

\noindent Analogous to the case of L\'evy flight, also for L\'evy walks and for the L\'evy-Lorentz gas one can define a characteristic length scale $\ell(T)$, which grows with time $T$, by considering the asymptotic scaling form of a PDF which depends on time $T$ and on the distance from the starting point $X$. In particular, for the L\'evy walk $\ell(T) \sim T^{\gamma}$, with $\gamma=1$ for $\alpha<1$ i.e. ballistic walkers, $\gamma=1/\alpha$ for $1<\alpha<2$ i.e. superdiffusive walkers, and $\gamma=1/2$ for $\alpha>2$ i.e. standard diffusive walkers. The case of $\alpha<1$ is not considered in our analysis because in this case the single big-jump principle does not apply, since it is not possible for a single jump to reach a distance much larger than the typical scale,  due to the fact that both the typical scale and the single jump follows a ballistic dynamics. The single big-jump principle is only valid for $\alpha>1$, and these are the cases we are interested in. Consider a one-dimensional L\'evy walk whose position after a time $t$ is denoted by $x(t)$, assuming that it starts from the origin $x(0)=0$. The process at each time can be divided into a discrete set of time intervals $[t_1, ..., t_n, ...]$ representing the times at which individual jumps occur, and the duration of each $n$-th jump $t_{n+1} - t_{n} = \delta t$ is extracted from the PDF \eqref{levywalk}. For $t_{n} \leq t \leq t_{n+1}$, the position $x(T)$ of the random walker at time $t$ is: 

\begin{equation}
x(t) = x(t_{n}) + \sign(v_{n+1}) \times v (T-t_{n})
\end{equation}

\noindent where $\sign(v_{n+1})$ is positive or negative with probability $1/2$. We define $P(X, T)$ as the probability density that, in a time interval $T$, the maximum value reached by the L\'evy walk is $X$. If $X$ is much larger than the typical length of the system ($\ell(T) \sim T^{1/\alpha}$ for $1<\alpha<2$ and $\ell(T) \sim T^{1/2}$ for $\alpha>2$) the single big-jump principle can be applied. Thus, the only significant contribution that leads the L\'evy walk maximum beyond $X$ is given by a single macroscopic jump of duration $X/v$. In the remaining time $T - X/v$ of the process small contributions occur that bring the process to values on the order of the typical length, so they can be neglected. Similarly to the previous case we consider the probability that the maximum value of $x(t)$ is larger than $X$. According to the big-jump principle we obtain:

\begin{align}
\label{probwalk}
\text{Prob}\left(\max_{t \in [0, T]} x(t) > X \right)&\sim \frac{1}{2} N_{eff} \times \text{Prob}(t > X/v) \\ \nonumber
&\sim  \frac{T-X/v}{2 \langle t \rangle} \int_{X/v}^{\infty} dt \ p(t)
\end{align}

\noindent Again the $1/2$ factor accounts for the fact that individual steps are equally likely to be drawn in the positive or negative direction. $\text{Prob}(t > X/v)$ is the probability that a step is larger than $X$, while $N_{eff}$ represents the number of steps (attempts to make the big jump) in a process where there is a single big jump of length $X\sim vT$. We have $N_{eff}=(T-X/v)/\langle t \rangle$
where $\langle t \rangle = \int dt' \ t' p(t')$ is the average duration of a single step and $T-X/v$ is the time available. Deriving with respect to $X$ Eq. \eqref{probwalk}, the PDF of the running maximum for the L\'evy walks, when $\alpha>1$, is:

\begin{align}
P(X, T) &\sim \frac{1}{2} \frac{1}{ v \langle t \rangle} \int_{X/v}^{\infty} dt \ p(t) + \frac{1}{2}\frac{T-X/v}{v \langle t \rangle} p(X/v) \\ \nonumber
&\sim \frac{1}{2 v \langle t \rangle} (v t_0)^{\alpha} X^{-\alpha} + \frac{T-X/v}{2v \langle t \rangle} \alpha (v t_0)^{\alpha} X^{-\alpha-1} \\ \nonumber
&\sim B_0(X, T) + B_1(X, T)
\end{align}

\noindent where we can recognize the exact asymptotic expression of the tail PDF for position $X$ at time $T$ of a L\'evy walk obtained from the single big-jump approach in \cite{BJ1}. In particular, we find the same pair of contributions $B_0(X,T)$ and $B_1(X, T)$, which in the single big-jump interpretation represent, respectively, the probability distribution that the single big jump starts at time $T-X/v$ and is partially completed in the interval $T$ and conversely the probability distribution that it starts at a smaller time than $T-X/v$ and it completes its step entirely in the time interval $T$. By rescaling the process with respect to the normalized variable $r=X/(vT)$ it is possible to rewrite the probability distribution $P(X, T)$ in terms of an infinite density function $I_{\alpha}(r)$, i.e., $P(X, T) \sim B_0(X, T) + B_1(X, T) \sim T^{-\alpha} I_{\alpha} (X/(vT))$, with:
\begin{equation}
\label{levywalkpdf}
I_{\alpha}(r) = \frac{(v t_0)^{\alpha}}{2 v \langle t \rangle} \left[\alpha r^{-\alpha-1} - (\alpha - 1) r^{-\alpha}\right]  
\end{equation}
Eq. \eqref{levywalkpdf} shows that, analogously to the case of the PDF of the position \cite{BJ1}, also for the EVS, the constant velocity of the steps naturally introduce a linearly growing scaling length in the shape of the tail which describes the rare events. We will show that the same behavior is present also in the L\'evy-Lorentz gas where the velocity in the steps is also constant. On the other hand Eq. \eqref{levyflightpdf} shows that for L\'evy flights the tail of the distribution is scale free, since in this case steps are instantaneous and they do not introduce any characteristic length growing with time. In Figure \ref{fig:walk} it is shown that the probability distribution of the running maximum $P(X, T)$ plotted against $X/(vT)$ converges excellently for large times to the result of Eq. \eqref{levywalkpdf}.

\section{Extreme Value Statistics of L\'evy-Lorentz Gases}
\label{section4}

\noindent L\'evy-Lorentz gases are one-dimensional arrays of scatterers distributed over a disordered lattice where each distance between two successive scatterers follows a probability distribution with a power tail. The distance $l$ between a pair of neighbours determines the structure of the random lattice. In particular we drawn the distances from a PDF with a lower cut-off $l_0$:

\begin{equation}
\label{levyglass}
p(l) = 
\begin{cases}
\frac{\alpha l_0^{\alpha}}{l^{1 +\alpha}} & l>l_0 \\
0 & l<l_0
\end{cases}
\end{equation}

\noindent It is natural to define a random walk on the L\'evy-Lorentz gas, just take a walker moving with constant velocity $v>0$ between two scatterers and choose a starting point on the lattice. Each time the walker hits a scatterer, it has a probability $\epsilon$ of being reflected reversing its motion and a probability $1-\epsilon$ of being transmitted and continuing the motion in the same direction to the next scatterer. For convenience, at time $t=0$ the walker starts at the scattering site in $x=0$ and the process is symmetric, i.e. $\varepsilon=1/2$. The PDF for the position of a walker at a generic time in a L\'evy-Lorentz gas has been studied  in \cite{LevyLorentzBarkai, LevyLorentzBeenakker}, and also its asymptotic behaviour in the tail limit has been obtained through the single big-jump principle \cite{BJ1, BJ3}. The PDF is characterised by a scaling length $\ell(T)$, i.e. the typical distance covered by the walker in a time $T$. As reported in \cite{LevyLorentz1, LevyLorentz2Univ, LevyLorentzBarkai}, for $\alpha < 1$ the walker has a superdiffusive dynamics with $\ell(T) \sim T^{\frac{1}{1+\alpha}}$, while for $\alpha > 1$ the motion is diffusive and $\ell(T) \sim T^{\frac{1}{2}}$. Now consider a one-dimensional L\'evy-Lorentz gas, and denote its position at time $t$ by $x(t)$. The random environment in which the stochastic process takes place can be described by a discrete set of positions $[l_1, ..., l_n, ...]$, representing the positions of the scatterers visited by the random walker. The distances between two consecutive scatterers $l_{n+1} - l_{n} = \delta l$ are drawn according to the PDF \eqref{levyglass}. Moreover, in this model the index of the lattice points over which the walker moves, i.e., $n(k)$, is also defined by a stochastic process. In particular, it is a stochastic process on the integers defined by the following jump rule:

\begin{figure}
\includegraphics[scale = 0.6]{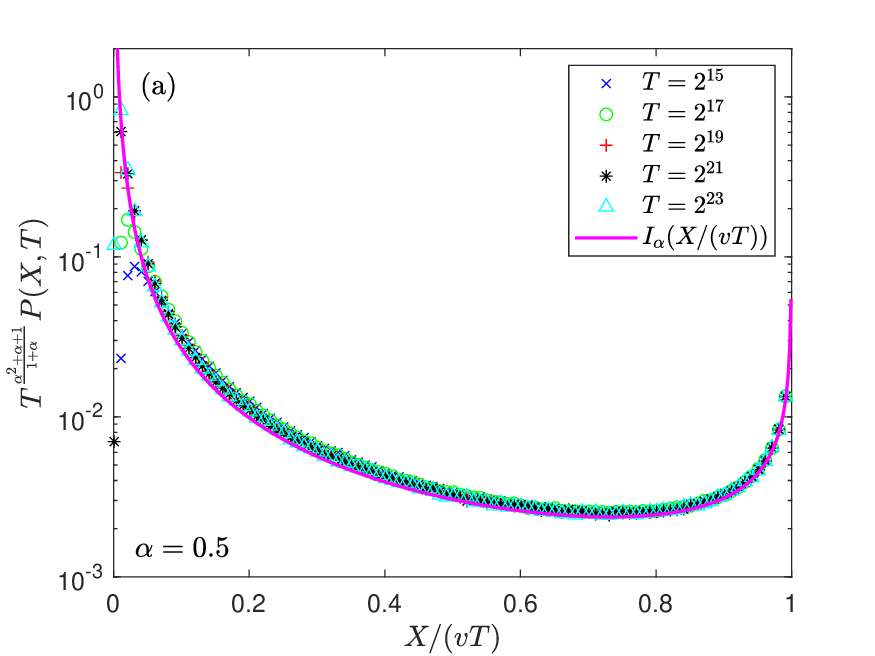}
\includegraphics[scale = 0.6]{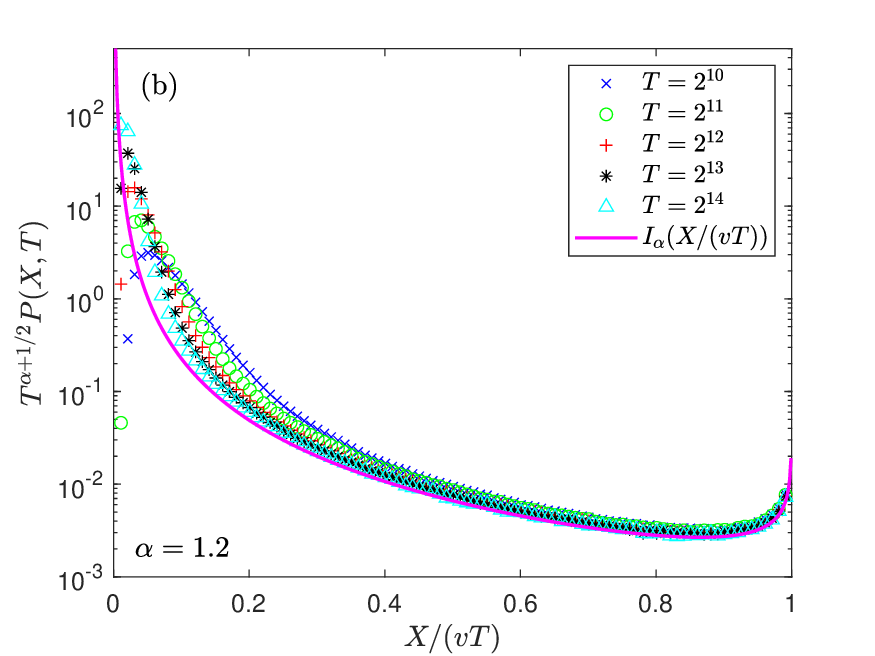}
\caption{Probability density function of running maximum $P(X, T)$ for L\'evy-Lorentz gases at different times $T$. (a) plots an example of a superdiffusive walker with $\ell(T) \sim T^{\frac{\alpha}{1+\alpha}}$ for $\alpha = 0.5$, while (b) plots an example of a standard diffusive walker with $\ell(T) \sim T^{1/2}$ for $\alpha=1.2$. In both simulations, $l_0=1, v=1$ were fixed and the probabilities were multiplied by a normalization factor $T^{\frac{\alpha^2+\alpha+1}{1+\alpha}}$ for (a) and by $T^{\frac{1}{2}+\alpha}$ for (b). The continuous pink line represents the asymptotic analytic prediction in terms of the rescaled function found in Eq. \eqref{levyglasspdf}. In this case a multiplicative constant has been
optimized for the estimate of the prefactor $\tau_0$ in order to reproduce numerical data.}
\label{fig:glass}
\end{figure}

\begin{equation}
\label{randindices}
n(k+1) = n(k) \pm 1
\end{equation}

\noindent where  the index increases by one with a probability $1/2$ and decreases with probability $1/2$. This stochastic process on the integers takes into account the random reflection and transmission events on each individual scatterer, keeping track of the path previously taken by the walker. The process can now be divided into a discrete set of times $[t_1, ..., t_k, ...]$, where the individual time steps are defined as $t_{k+1} = t_k + \big| \frac{l_{n(k+1)}-l_{n(k)}}{v} \big|$,  where $l_{n(k)}$ is the position of the walker at the $k$-th step at time $t_{k}$. Considering the combination of the stochastic process for extracting the random lattice distances \eqref{levyglass} and the stochastic process for the current position of the lattice indices \eqref{randindices}, it is possible to write the equation for the position $x(t)$ of a random walker in a L\'evy-Lorentz gas, assuming that $t_{k} \leq t \leq t_{k+1}$ (for a more rigorous description see \cite{Bianchi2016}):

\begin{equation}
x(t) = l_{n(k)} + [n(k+1) - n(k)] \times v (t-t_{k})
\end{equation}

\noindent where $l_{n(k)}$ is the position of the walker on the lattice at time $t_{k}$ and the difference $n(k+1)-n(k)$ defines the sign of the spatial increment $r(t) = v(t-t_{k})$, which has probability $1/2$ to be negative and probability $1/2$ to be positive. The object of interest is the probability distribution $P(X, T)$ that the extreme value of the position of the walker on the L\'evy-Lorentz gas reaches a maximum $X$ within a time interval $T$. Assuming that $X$ is much larger than the typical length of the system $X \gg \ell(T)$ for both $\alpha<1$ and $\alpha>1$, the single big-jump principle can again be applied. So, the only significant contribution that takes the maximum value of the position beyond $X$ occurs when the walker crosses a scattering point that is separated from its next by a distance greater than $X$. After crossing this long gap, the motion can be treated as deterministic, since the borders of the gap act as perfectly reflecting walls on time scales of order $T$. The motion of the walker before the big jump can be neglected since it is of order of the typical length scale $\ell(T)$. During the time interval $T$ the walker visits, in the remaining time $T - X/v$, an effective number of crossing sites $N_{eff}$. This number has already been estimated in \cite{LevyLorentz2Univ, LevyLorentzBarkai, LevyLorentzBeenakker}, and it has been shown that for sufficiently long times $N_{eff} \sim ((T - X/v) / \tau_0)^{\frac{\tilde \alpha}{1+\tilde \alpha}}$ 
with $\tilde{\alpha} = \alpha$ for $\alpha<1$ and  $\tilde{\alpha} = 1$ for $\alpha>1$, while $\tau_0$ is a suitable time constant. At this point, applying the single big-jump principle, the random walker has $N_{eff}$ attempts before it performs a single big jump that carries it to a scatterer located at a distance greater than $X$ from the previous one in a time interval $X/v$, and therefore the probability that the running maximum is greater than $X$ in the limit $X \gg \ell(T)$ is:

\begin{align}
\label{probglass}
\text{Prob}\left(\max_{t \in [0, T]} x(t) > X \right)&\sim \frac{1}{2} N_{eff} \times \text{Prob}(l > X) \\ \nonumber
&\sim  
\frac{1}{2}  \left( \frac{T-X/v}{\tau_0} \right) ^{\frac{\tilde{\alpha}}{1+\tilde{\alpha}}} \int_{X}^{\infty} dl \ p(l)
\end{align}

\noindent The factor $1/2$ takes into account the probability to be transmitted. Finally, deriving with respect to $X$ the probability \eqref{probglass}, the PDF of the running maximum for the L\'evy-Lorentz gas is:

\begin{widetext}
\begin{align}
P(X, T) &\sim \frac{1}{2} \left[ \frac{1}{v \tau_0^{\frac{\tilde{\alpha}}{1+\tilde{\alpha}}}} \frac{\tilde{\alpha}}{1+\tilde{\alpha}} (T-X/v)^{\frac{\tilde{\alpha}}{1+\tilde{\alpha}} -1}\int_X^{\infty} dl p(l) + \left(\frac{T-X/v}{\tau_0} \right)^{\frac{\tilde{\alpha}}{1+\tilde{\alpha}}} p(X)\right] \\ \nonumber
&\sim \frac{l_0^{\alpha}}{2\tau_0^{\frac{\tilde{\alpha}}{1+\tilde{\alpha}}}} (T - X/v)^{\frac{\tilde{\alpha}}{1+\tilde{\alpha}}} X^{-\alpha} \left[\frac{1}{v} \frac{\tilde{\alpha}}{1+\tilde{\alpha}}\frac{1}{(T-X/v)} + \frac{\alpha}{X} \right] \\ \nonumber
&\sim B_0(X, T) + B_1 (X, T) 
\end{align}
\end{widetext}

\noindent  Interestingly, the PDF of the running maximum of a walker in a L\'evy-Lorentz gas has two contributions, $B_0(X,T)$ and $B_1(X,T)$. $B_0(X,T)$ comes from the processes where the walker makes the big jump at time $T-X/v$ and still travels the long gap at time $T$ without hitting the last scatterer. In $B_1(X,T)$ the walker hits the last scatterer of the long gap before time $T$ and continues to be reflected in the gap for the remaining time. Therefore, in the L\'evy-Lorentz gas, the PDF of the EVS differs significantly from the PDF of the position \cite{BJ1}, where it is necessary to keep track of all the reflections that occur once the random walker enters the long gap, and this gives rise to an infinite sum of contributions, one for each reflection. On the other hand, in the case of the running maximum, once the extreme is reached, the value of the maximum is simply the length of the long gap, independent of the reflection dynamics that determine the final position of the walker. In other words, unlike the L\'evy flight and the L\'evy walk, the quenched topology along which this walker moves induces memory effects in its dynamics, moving it away from the maximum once it has been reached for the first time.   Now, rescaling the process with respect to the normalized variable $r=X/(vT)$, the PDF of the maximum can be rewritten in terms of a function $I_{\alpha}(r)$:

\begin{widetext}
\begin{equation}
P(X,T) = 
\begin{cases}
T^{-\frac{\alpha^2 + \alpha +1}{1+\alpha}} I_{\alpha} \left(\frac{X}{vT}\right)  & \alpha<1 \\
T^{-\frac{1}{2} - \alpha} I_{\alpha} \left(\frac{X}{vT}\right)  & \alpha>1 
\end{cases}
\end{equation}

\noindent where the asymptotic expression for the rescaled function $I_{\alpha}(r)$ of the running maximum in the L\'evy-Lorentz gas is:

\begin{equation}
\label{levyglasspdf}
I_{\alpha}(r) = 
\begin{cases}
\frac{l_0^{\alpha}}{2v^{\alpha+1} \tau_0^{\frac{\alpha}{1+\alpha}}} \alpha  (1-r)^{\frac{\alpha}{1+\alpha}} r^{-\alpha}\left[ \frac{1}{1+\alpha}\frac{1}{1-r} + \frac{1}{r}\right] & \alpha<1 \\
\frac{l_0^{\alpha}}{2v^{\alpha+1} \tau_0^{\frac{1}{2}}} (1-r)^{\frac{1}{2}} r^{-\alpha}\left[ \frac{1}{2}\frac{1}{1-r} + \frac{\alpha}{r}\right] & \alpha>1
\end{cases}
\end{equation}
\end{widetext}

\noindent In Figure \ref{fig:glass} we show that the probability distribution of the running maximum $P(X, T)$ plotted against $X/(vT)$ converges successfully for large times to the result of Eq. \eqref{levyglasspdf}, confirming the validity of the single big-jump approach;

\section{Conclusions}

The rare events in the EVS of one-dimensional L\'evy flights, L\'evy walks, and L\'evy-Lorentz gases, all of which involve steps drawn from power-tailed probability distributions, have been studied through an analysis based on the single big-jump principle, a powerful tool in the study of rare events in subexponential tailed probability distributions.
For these three stochastic jump processes we derive the tail of the probability density function of the running maximum of the position. \\
Our results extend very recent previous results on L\'evy flights \cite{Klinger2023} and present the more complex cases of  L\'evy walks and L\'evy-Lorentz gases. In particular,  the L\'evy-Lorentz gases are much more difficult to study due to the disordered topology of the lattice on which the process takes place. 
For L\'evy flights and L\'evy walks, the big-jump principle approach implies that after the big jump the walker remain fixed at its maximum so that the rare event statistics of the extreme values and of the position coincide. For the L\'evy-Lorentz gas, the topology of the disordered lattice affects the walker dynamics and influences the statistics of rare events. In particular the value of the maximum is determined by the length of the big jump but the position is determined by the reflection dynamics that occur once the random walker enters the long gap \cite{BJ1}, so that the rare event statistics of the extreme values and of the position are different. All the analytical predictions obtained by this heuristic argument are validated by extensive numerical simulations, showing excellent agreement for both superdiffusive and standard diffusive cases. This suggests that the estimate is essentially correct, and this can open a path to a rigorous derivation. Throughout the analysis, the single big-jump principle proves to be a robust approach for estimating the probability distribution of extreme values in various L\'evy processes. The results highlight the applicability of the principle for understanding rare events in different scenarios, but we expect such estimates to
hold for the EVS of a larger class of sub-exponential jump processes, providing a new useful approach in the field of EVS and record statistics;

\section{Acknowledgement}

\noindent This research was granted by University of Parma through the action Bando di Ateneo 2022 per la ricerca co-funded by MUR-Italian Ministry of Universities and Research - D.M. 737/2021 - PNR - PNRR - NextGenerationEU (project  'Collective and self-organised dynamics: kinetic and network approaches');

\appendix

\section{Running Maximum PDF and Survival Probability}
\label{appendixA}

In this section we show that the formula \eqref{levyflightpdf} in section \ref{section2} can be easily derived from the survival probability approach used in \cite{Klinger2023, MajumdarRecords}, exploiting the equivalence with the result obtained through the single big-jump principle. Consider a jump process having $n$-th steps $x(n)$ subject to Markov's rule \eqref{markov}, with $r(n)$ IID random variables drawn from a symmetric and continuous probability distribution $p(r)$. Define the maximum value of the jump process $M(n) = \max (x(0), x(1), ..., x(n))$  and assume it starts from the origin, i.e., $x(0)=0$. The object of interest is the probability that the maximum $M(n)$ is less than a threshold value $X$, i.e.:

\begin{equation}
\text{Prob}(M(n) \leq X) = \text{Prob}(\{ x(i) \leq X \ \  \forall i = 1,..., n\})
\end{equation}

\noindent Now, defining a new random variable $z(n) = X-x(n)$, the shift by $X$ of the equation \eqref{markov} yields the jump rule for the process $z(n)$, which starts at $z(0) = X$ and evolves in $n$ steps between $X$ and $M(n)-X$, and for this new process the probability that $M(n)$ does not exceed $X$ becomes the survival probability $q(X, n)$ that $z(n)$, that starts at $X$ and walks along $n$ steps, does not cross zero, i.e. $\text{Prob}(M(n) \leq X) = q(X,n)$. As reported in \cite{MajumdarRecords}, the survival probability on zero of a jump process with individual steps extracted from a power-tailed distribution $p(r) \sim r^{-1-\alpha}$ with $0<\alpha<2$ that starts from an initial position $X$ after $n$ steps in the limit of $X, n \rightarrow \infty$ with $X/n^{1/\alpha}$ fixed is:

\begin{equation}
\label{survival}
q(X, n) = 1 - \frac{n}{\pi} \Gamma(\alpha) \sin \left( \frac{\pi \alpha}{2} \right) \left( \frac{b_{\alpha}}{X} \right)^{\alpha}
\end{equation} 

\noindent Where $b_{\alpha}$ represents the typical length scale of the jump process with respect to the tail of $p(r)$, i.e., is the coefficient of the second order term in the small $k$ expansion of the PDF $\tilde{p}(k)$:

\begin{equation}
\label{smallk}
\tilde{p}(k) \underset{k \rightarrow 0}{=} 
 1 - (b_{\alpha} |k|)^{\alpha} + o(|k|^{\alpha})
\end{equation}

\noindent Now, we focus in determining the typical length scale $b_{\alpha}$. In the case of $\alpha \geq 2$ the typical length associated with the jump is related to the analytic expression of the second moment of the distribution $\langle r^2 \rangle$ which has finite value, and in the case of $\alpha=2$ it is $b_2 = \langle r^2 \rangle/\sqrt{2}$. In the case of $\alpha<2$ the second moment of the distribution $p(r)$ is not finite, and thus the expansion in small $k$ is not trivial. In the case of $\alpha<2$, especially in case of $1<\alpha<2$ (the same calculation can be repeated for $\alpha<1$)  the first moment of the distribution $\langle r \rangle$ is finite, while the second moment $\langle r^2 \rangle$ diverges. This means that the divergence in the expansion is contained in the second derivative of $\tilde{p}(k)$ in $k$, and it holds:

\begin{align}
\frac{\partial^2}{\partial k^2} \tilde{p}(k) &= \frac{\partial^2}{\partial k^2} \int_{-\infty}^{+\infty} dr e^{ikr} p(r) \\ \nonumber
&= - \int_{-\infty}^{+\infty} dr (\cos(kr)+ i \sin(kr))p(r) r^2 \\ \nonumber
&= - \int_{0}^{+\infty} dr \cos(kr) p(r) r^2
\end{align}

\noindent In the last line we use that $p(r)$ is symmetric and defined only for positive values of $r$. Now inserting the probability distribution \eqref{levyflight}, and by making the change of variables $r' = kr$, the divergent term in $k$ can be taken out of the integral, and in the limit of $k \rightarrow 0$ the integral converges to a finite value:

\begin{align}
\frac{\partial^2}{\partial k^2} \tilde{p}(k) &= - \alpha r_0^{\alpha} |k|^{\alpha-2} \int_{k r_0}^{\infty} dr' r'^2 \frac{\cos(r')}{r'^{1+\alpha}} \\ \nonumber
&\underset{k \rightarrow 0}{=} \alpha r_0^{\alpha} |k|^{\alpha-2} \cos\left(\frac{\alpha \pi}{2}\right) \Gamma(2-\alpha)
\end{align}

\noindent Now we can return to the small $k$ expansion \eqref{smallk} just integrating this result twice, imposing normalization on the zero-order term in $k$ and $\langle r \rangle = 0$ by symmetry:

\begin{align}
\tilde{p}(k) &= 1 - r_0^{\alpha} \cos\left( \frac{\alpha\pi}{2} \right) \Gamma(1-\alpha) |k|^{\alpha} + o( |k|^{\alpha})\\ \nonumber
&= 1 - (b_{\alpha} |k|)^{\alpha} +o( |k|^{\alpha})
\end{align}

\noindent One can obtain the same result for $b_{\alpha}$ even in the case of $\alpha<1$ just studying the divergence of the first derivative of $\tilde{p}(k)$ and repeating the same calculations. Now, the PDF that the maximum is greater than $X$ after $n$ steps $P(X, n)$ becomes, substituting the expression derived for $b_{\alpha}$ in \eqref{survival}:

\begin{align}
P(X, n) &= \frac{d}{d X} (1-q(X, n)) \\ \nonumber
&= \frac{\alpha n}{\pi} \Gamma(\alpha) \sin\left( \frac{\alpha \pi}{2} \right)r_0^{\alpha} \cos\left( \frac{\alpha\pi}{2} \right) \frac{\Gamma(1-\alpha)}{X^{\alpha+1}} \\ \nonumber
&= \frac{1}{2} n \left( \frac{\alpha r_0^{\alpha}}{X^{\alpha+1}} \right) \\ \nonumber
&= \frac{1}{2} n p(X)
\end{align}

\noindent Where we use between the second and the third lines the sine duplication formula $\sin(2x)=2\cos(x)\sin(x)$ and the property of Gamma functions $\Gamma(1-x)\Gamma(x)=\frac{\pi}{\sin (\pi x)}$ to simplify some factors. Finally, we show that the survival probability approach in the limit of large $X$ coincides with the single big-jump approach for the case of L\'evy flights. This means that the linear dependence of $P(X,n)$ on $n$ and $p(X)$ can be interpreted in the sense of single big jump: the walker has exactly $n$ attempts to make a macroscopic jump that brings it close to $X$. The same kind of heuristic argument can be repeated for L\'evy walks and for L\'evy-Lorentz gases discussed in sections \ref{section3} and \ref{section4} just replacing $n$ with $N_{eff}$;

\bibliography{BJ_biblio}

\end{document}